# An Analysis of Anomaly Cancellation
# for Theories in D=10

Andrea Antonelli [1,2]

[1]Department of Physics, Faculty of Science,
University of Tokyo, Bunkyo-ku, Tokyo 133-0022, Japan
[2]Department of Physics, King's College London,
The Strand, London WC2R 2LS, United Kingdom

## Abstract

We prove that the swampland for D=10 $\mathcal{N} = 1$ SUGRA coupled to D=10 $\mathcal{N} = 1$ SYM is only populated by $U(1)^{496}$ and $E_8 \times U(1)^{248}$. With this goal in mind, we review the anomalies for classical and exceptional groups, retrieving trace identities up to the sixth power of the curvature for each class of groups. We expand this idea for low-dimensional groups, for which the trace of the sixth power is known to factorize, and we retrieve such factorization. We obtain the total anomaly polynomials for individual low dimensional groups and combinations of them and finally we investigate their non-factorization, in such a way that $U(1)^{496}$ and $E_8 \times U(1)^{248}$ are non-trivially shown to be the only anomaly-free theories allowed in D=10. Using the method developed for checking the factorization of gauge theories, we retrieve the Green-Schwarz terms for the two theories populating the swampland.

# Contents

# 1. Introduction

## 1.1 Motivations behind this paper and organization of the material

In recent years, following the work of Vafa, the "swampland program" has aimed at giving a boundary to the swampland of the effective theories that are not fully embedded in theories of quantum gravity [11,12]. One important tool at hand for physicists is the concept of quantum anomaly, and consequently anomalies have been studied in detail.

In D=10 it has long been known [3] that $SO(32)$, $E_8 \times E_8$, $U(1)^{496}$ and $E_8 \times U(1)^{248}$ are theories where quantum anomalies are cancelled. The former two are also low energy limits of string theories and do not therefore pertain to the swampland.

$U(1)^{496}$ and $E_8 \times U(1)^{248}$, conversely, do live in the swampland and it was not known until recently whether or not they were also low energy limits of string theories. Following the work of Fiol [10] and the joint work of Adams, DeWolfe and Taylor [4] it was shown that there are no theories of gravity that can be coupled to them, thus confining both theories to the swampland, without possibilities to be upgraded to string theories. However, there is still one unanswered question: are $SO(32)$, $E_8 \times E_8$, $U(1)^{496}$ and $E_8 \times U(1)^{248}$ the only theories where the anomalies are cancelled?

Hints for this statement can be traced back in the literature, mostly following the original statement found in chapter 13 of [3], but no proof has explicitly been carried out. Therefore answering this question might help understand the structure of the swampland in D=10.

In this paper it is shown that indeed there are no other theories in the swampland in D=10: in the process of proving, we devise some non-trivial machinery that is worthwhile presenting and that might be useful in anomaly cancellation in other dimensions; also, we deal with low-dimensional theories that are often overlooked in the analysis of anomaly cancellation.

The reader is encouraged to read the first part of this paper, as there most of the material and notation is presented: more in detail, the Green-Schwarz Mechanism is revisited in the notation of Bilal [5] and some important points for the upcoming sections are explained. Plus, proof with appropriate references are given for the non-factorization of the trace of the sixth power of the curvature in classical and exceptional groups, following the Cartan classification of simple Lie groups.

The expert reader might want to skip the classification and non-factorization of classical and exceptional groups and dive into the last part involving low-dimensional groups. There, the concept of factorization is treated in detail and it is shown why anomalies of individual groups are carried along when combining them and specifically what non-canceling terms of the polynomial contain the information about the anomaly of the theory.

This paper might be of interest also to readers looking for the factorization or non-factorization of the trace of the sixth power of the curvature in all classical, exceptional and low-dimensional groups, although a fairly more rigorous treatment for most (but not all) of them is found in [6] and [8].

## 1.2 The Green-Schwarz Mechanism

It is instructive at this point to review the mechanism devised in [2] that allows anomaly cancellation. In doing so, the reader will be reminded of numerous concepts that will be fundamental for understanding the upcoming sections. Moreover, several of the ideas presented herein will be useful when the Green-Schwarz mechanism is used for non-trivial groups in the final sections.

We will begin saying that all the information about the anomaly of a system can be encapsulated in the total anomaly polynomial [5]: in principle, then, such a polynomial can vanish if a counterterm depending on the choice of a particular gauge transformation is added. If this happens, the gauge theory becomes non-anomalous (or, equivalently, anomaly-free).

The Green-Schwarz mechanism does exactly this; the groundbreaking idea was to devise an "inflow" method to construct a counterterm and then recast it in terms of a 12-form polynomial, in such a way to be able to add it to the total anomaly polynomial and cancel it.

We will now explain the method in much more detail referring to a ten dimensional gauge group G that is a $\mathcal{N}$ =1 Supergravity coupled to a $\mathcal{N}$ =1 Super Yang-Mills theory. As a remainder, the matter content of SUGRA is a multiplet comprising a positive chirality Majorana-Weyl gravitino and a negative chirality spin ½ fermion. Moreover, SUGRA contains the graviton, a scalar renamed dilaton and a two-form $B$.

On the other hand, the Super Yang Mills contains a multiplet of gauge fields $A_\mu^\alpha$ and gauginos $\chi^\alpha$ that live in the same adjoint representation [5].

The requirement for a consistent coupling is given by the following expression for the field strength:

$$H = dB - Q_3(A, F) + \beta Q_3(\omega, R) \equiv dB - Q_3^{YM} + \beta Q_3^L$$

where $Q_3(A, F)$ is the gauge Chern-Simons form and $Q_3(\omega, R)$ is the gravitational one [5]. The invariance of H when acted upon with a gauge and Lorentz transformation is given as follows:

$$(\delta^{gauge} + \delta^L)B = Q_2^{YM,1} + \beta Q_2^{L,1}$$

Given the above ingredients, it is now possible to choose a counterterm with an appropriate 8-form $X_8$ retrieved from the characteristic classes:

$$\Delta\Gamma = \int B \wedge X_8 \quad such\ that$$

$$(\delta^{gauge} + \delta^L)X_8 = 0 \quad and$$

$$X_8 = dX_7$$

When we act upon the counterterm with a gauge and Lorentz transformation we obtain:

$$(\delta^{gauge} + \delta^L)\Delta\Gamma = \int (Q_2^{YM,1} + \beta Q_2^{L,1}) \wedge X_8 = \int (Q_2^{YM,1} + \beta Q_2^{L,1}) \wedge dX_7$$

$$= -\int (\delta\, Q_3^{YM} - \beta\delta Q_3^{I})\wedge X_7 \tag{1.2.1}$$

Using the descent equations, (1.2.1) can be recast in terms of 12-form polynomial:

$$\Delta\hat{I}_{12} = (trF^2 - \beta\, trR^2)\wedge X_8$$

The 12-form polynomial tells us that the total anomaly polynomial is cancelled by adding the counterterm *if and only if it does not vanish and takes a factorized form*. This idea is so important that we can safely say that the rest of this report will be centered on the concept of factorization: specifically we will be interested in if and how the total anomaly polynomial factorizes and therefore vanishes upon adding a counterterm $\Delta\Gamma$.

Consider again the gauge group G. Its matter content is such that the various contributions to the anomaly are [5]:

$$I_{12}^{gravitino} = \frac{1}{64(2\pi)^5}\left(-\frac{11}{126}trR^6 + \frac{5}{96}trR^4trR^2 - \frac{7}{1152}(trR^2)^3\right) \tag{1.2.2.a}$$

$$I_{12}^{spin\frac{1}{2}} = \frac{1}{64(2\pi)^5}(tr_{\mathcal{R}}1)\left(\frac{1}{5670}trR^6 + \frac{1}{4320}trR^4trR^2 + \frac{1}{10368}(trR^2)^3\right)$$
$$- \frac{1}{32(2\pi)^5}(tr_{\mathcal{R}}F^2)\left(\frac{1}{360}trR^4 + \frac{1}{288}(trR^2)^2\right) + \frac{1}{1152(2\pi)^5}(tr_{\mathcal{R}}F^4)trR^2$$
$$- \frac{1}{720(2\pi)^5}(tr_{\mathcal{R}}F^6) \tag{1.2.2.b}$$

The total anomaly polynomial is given by the contribution of all the components of the matter content and can be expressed as follows:

$$I_{12}^{total} = I_{12}^{gravitino} - I_{12}^{spin\frac{1}{2}}|_{\mathcal{R}=1} + I_{12}^{spin\frac{1}{2}}|_{\mathcal{R}=adj}$$

$$I_{12}^{total} = \frac{1}{64(2\pi)^5}\left[\frac{n-496}{5670}trR^6 + \frac{n+224}{4320}trR^4trR^2\right.$$
$$\left. + \frac{n-64}{10368}(trR^2)^3\right]$$
$$- \frac{1}{32(2\pi)^5}(TrF^2)\left(\frac{1}{360}trR^4 + \frac{1}{288}(trR^2)^2\right) + \frac{1}{1152(2\pi)^5}(TrF^4)trR^2$$
$$- \frac{1}{720(2\pi)^5}(TrF^6) \tag{1.2.3}$$

Where $Tr$ and $tr$ represents the trace in the adjoint and fundamental representations respectively and $n$ is the dimensionality of the gauge group.

From (1.2.3) two conditions emerge in order for the anomaly to be cancelled. Firstly, the number of the dimensions of the gauge group must be 496 in order for $trR^6$ to disappear. A second, equally important condition is given by the factorization of the last term $TrF^6$. Indeed, if this does not factorize the Green-Schwarz mechanism cannot be used. This idea will be explained in detail later: for the sake of completeness, however, it is now anticipated that the total anomaly polynomial must factorize in such a way that only 8-form and 4-form terms appear, since such is the coupling of the electric and magnetic currents. Clearly, $TrF^6$ is a 12-form and does not pertain to either factor: factorizing this term in 8-forms and 4-forms allows recovering the shape of the coupling of currents, thus giving us a chance to cancel the anomaly via the Green-Schwarz mechanism.

In the next paragraph we explore further the latter condition and we shall see how (1.2.3) works for two choices of the gauge group G.

## 1.3 Anomaly Cancellation in $SO(32)$ and $E_8 x E_8$

Let us first focus on G $\equiv SO(32)$: such a Lie group has 496 generators, so its dimensions are just enough to cancel the first term of (1.2.3). The remaining problem is related to the trace of the sixth order of the curvature $F$, which should factorize. We will now anticipate some useful relations between the traces of the adjoint and those of the fundamental representations. The proof will be found in chapter 2, which deals with the more general $SO(n)$ group.

$$TrF^2 = 30 trF^2$$
$$TrF^4 = 24\ trF^4 + 3(trF^2)^2$$
$$TrF^6 = 15\ TrF^2\ TrF^4 \tag{1.3.1}$$

The reason why such trace identities are looked for is simple: in D=10 the SYM contains vector gauge fields and Majorana-Weyl spinors that live in the adjoint representation. Moreover the anomaly is found to be [3] proportional to a term $T$ constructed out of the trace of the elements of the gauge algebra $t$ that live in the adjoint representation. Such elements of the gauge algebra can sometimes be daunting to evaluate and for this reason relations between traces in the adjoint and fundamental representations are looked for: it is in general easier to deal with generators of the gauge algebra than with its elements.

Coming back to the point of the discussion, the relations (1.3.1) are inserted in the total anomaly polynomial to obtain a factorized form:

$$I_{12}^{total} = \frac{1}{384(2\pi)^5}(trR^2 - trF^2)\left[trR^4 + \frac{1}{4}(trR^2)^2 - trR^2 trF^2 + 8trF^4\right]$$

In order to cancel this it is clear that the counterterm $\Delta\Gamma$ must be constructed from the following chern 8-form:

$$X_8 = \frac{1}{384(2\pi)^5}\left[trR^4 + \frac{1}{4}(trR^2)^2 - trR^2 trF^2 + 8trF^4\right]$$

$$\Delta\hat{I}_{12} = \frac{1}{384(2\pi)^5}(trF^2 - trR^2)\left[trR^4 + \frac{1}{4}(trR^2)^2 - trR^2 trF^2 + 8trF^4\right]$$

The counterterm is now nothing more than the opposite of the total anomaly polynomial so that when the two are added, the anomaly vanishes via G-S mechanism:

$$\Delta \hat{I}_{12} + I_{12}^{total} = 0$$

A second candidate for a consistent theory is $E_8 \times E_8$. The dimensions of $E_8$ is 248, therefore the dimensions of the combination are consistent with the first term of the anomaly polynomial. Again trace identities can be retrieved [5]:

$$TrF^4 = 9\,(trF_1^2)^2 + 9\,(trF_2^2)^2, \ \ TrF^6 = \frac{75}{20}(trF_1^2)^3 + \frac{75}{20}(trF_2^2)^3$$

where the indices are used to distinguish between the curvatures of the two groups $E_8$ involved. The total anomaly polynomial again factorizes:

$$I_{12}^{total} = \frac{1}{384(2\pi)^5}\big(trR^2 - trF_1^2 - trF_2{}^2\big)$$
$$\left[ trR^4 + \frac{1}{4}(trR^2)^2 - trR^2\big(trF_1{}^2 + trF_2{}^2\big) - 2trF_1{}^2trF_2{}^2 + 2(trF_1^2)^2 + 2(trF_2^2)^2 \right]$$

The Green Schwarz term is then chosen to be:

$$X_8 = \frac{1}{384(2\pi)^5}\Big[ trR^4 + \frac{1}{4}(trR^2)^2 - trR^2\big(trF_1{}^2 + trF_2{}^2\big) - 2trF_1{}^2trF_2{}^2$$
$$+ 2(trF_1^2)^2 + 2(trF_2^2)^2 \Big]$$

So that, following the same reasoning as above, the total anomaly polynomial cancels.

It turns out, to conclude, that there are two more groups where the anomaly is cancelled, namely $U(1)^{496}$ and $E_8 \times U(1)^{248}$ [3]. These groups, however, do not give rise to string theories [4,10]; both groups will be treated conceptually in section 2.4.5 and algebraically in the Appendix.

## 2. Anomaly cancellation in Lie groups with 496 generators

### 2.1 Aim of the paper

It is often said that four groups cancel the anomaly: $SO(32)$, $E_8 \times E_8$, $U(1)^{496}$ and $E_8 \times U(1)^{248}$. For the first two explicit calculations have been carried out in the Introduction section, for the latter two it is recommended to read the Appendix of this paper.

Nonetheless the set of Lie groups with 496 dimensions contains many more groups. The situation looks then like the following:

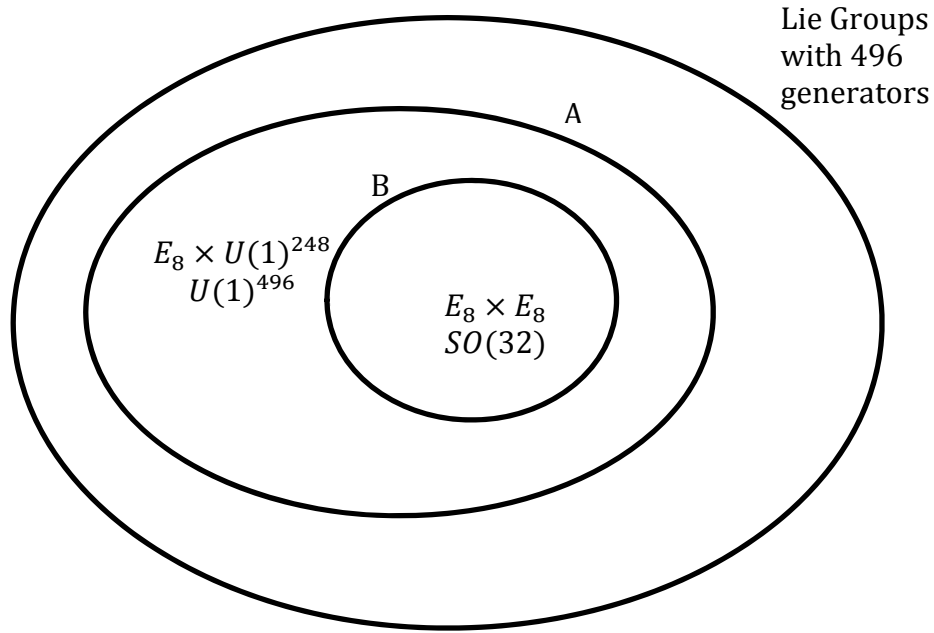

Fig 1. Set of all the Lie Groups in 496 dimensions and its internal classification.

Where the set A contains the theories in which the anomaly is cancelled and the subset B of theories embedded in a theory of quantum gravity.

This paper aims at checking that the above-mentioned quartet of groups contains all the gauge groups for which the anomaly is cancelled and therefore that |A|=4.

## 2.2 Cartan Classification of Simple Lie Groups

The Cartan classification of Lie groups into classical and exceptional groups comes in very handy for our purposes. All the Lie groups can be reduced to 4 classical and 5 exceptional groups [6]:

| Classical Groups | Exceptional Groups |
|---|---|
| $A_n$ ($n \geq 1$) compact<br>$B_n$ ($n \geq 2$) compact<br>$C_n$ ($n \geq 3$) compact<br>$D_n$ ($n \geq 4$) compact | $E_6$<br>$E_7$<br>$E_8$<br>$F_4$<br>$G_2$ |

Table 1. Cartan Classification of Simple Lie Groups.

The main idea is to check that gauge theories with 496 dimensions constructed out of classical groups, exceptional groups and mixtures of the two (with the exception of $E_8 \times E_8$) inevitably carry anomalies that cannot be canceled. We will see that such statement is intrinsically related to the non-factorization of the trace in the adjoint representation of the sixth order curvature $TrF^6$.

Once the classical and exceptional groups are found to be anomalous, it is obtained as a corollary that no anomaly-free theories can be retrieved coupling any of the group to $U(1)^n$ for a particular choice of $n$ (see section 3.1), again the only exception being the theory constructed via $E_8$, e.g. $E_8 \times U(1)^{248}$.

In the next paragraph we will show how the anomaly is preserved in some representatives of the classical groups, that nonetheless carry as much information as the above-mentioned $A_n, B_n, C_n, D_n$. In doing so, we will derive some useful trace identities that are crucial in our study of non-trivial groups in the last section.

## 2.3 Anomalies in Classical Groups

Throughout this chapter we assume that $n$ is sufficiently large, e.g. $n \geq 5$ for $A_n$, $n \geq 4$ for $D_n$ and $n \geq 3$ for $B_n$, and $C_n$. It will be soon explained that, for large $n$, $TrF^6$ does not factorize, whereas it does when low dimensional classical groups are considered. The latter case is thoroughly discussed in section 2.5.

### 2.3.1 $SO(n)$

To tackle this class of groups we need to find the relation between the trace of the adjoint representation and the one in the fundamental and demonstrate for which groups $TrF^6$ can or cannot factorize.

The starting point of the discussion is the action of the group: $SO(n)$ indeed acts on an anti-symmetric carrier space denoted by $v_{kl}$ that, being anti-symmetric, satisfies $v_{kl} = -v_{lk}$.

The transformation operated by the action of the gauge group is [7]:

$$v_{kl} \rightarrow v_{kl}^1 = (\Omega v)_{kl} \quad \text{with}$$

$$(\Omega v)_{kl} = \sum_{m<n} \Omega_{kl}^{mn} v_{mn} \tag{2.3.1.1}$$

The indices $k$ and $l$ run over the dimension of the group, e.g $k,l$=1,...., ½ $n(n-1)$, where we rename $N = \frac{1}{2}n(n-1)$ for future reference.

What has been found so far is an action via an adjoint representation of the gauge group. It is however possible to rewrite the adjoint transformation in terms of the fundamental representation $O$, so to link adjoint and fundamental in a first, unpolished, relation. This is done as follows (confront appendix F of [7]):

$$(\Omega v)_{kl} = \sum_{k'l'} O_{k'}^{k'} O_{l'}^{l'} v_{k'l'} = 2 \sum_{k'<l'} O_{k'}^{k'} O_{l'}^{l'} v_{k'l'} \tag{2.3.1.2}$$

By comparing (2.3.1.1) and (2.3.1.2) a relation between the action of the adjoint and fundamental representations is found:

$$\Omega_{kl}^{mn} = O_k^m O_l^n - O_k^n O_l^m \tag{2.3.1.3}$$

where the minus sign is due to the asymmetry of the carrier space and plays a crucial role in determining $SO(32)$ as the only possible consistent theory for $G = SO(n)$. Let us see how.

The relation (2.3.1.3) has been worked out to the point where the vector space does not appear anymore. Since we eventually want a relation between $Tr$ and $tr$ we calculate the former at this very stage:

$$Tr\Omega = \sum_{k<l} \Omega_{kl}^{kl} = \frac{1}{2}\sum_{k,l} \Omega_{kl}^{kl}$$

$$Tr\Omega = \frac{1}{2}\sum_{k,l=1}^{N}\left( O_k^k O_l^l - O_k^l O_l^k \right) = \frac{1}{2}\left[(trO)^2 - tr(O^2)\right] \qquad (2.3.1.4)$$

With equation (2.3.1.4) the trace of the adjoint has been related to the trace of the fundamental representation. However, the action as the argument for both is not as useful as a general element of the Lie Algebra: as far as anomalies are concerned, indeed, the trace is usually calculated over the curvature $\mathcal{F}$, which is an element of the Lie algebra.

The easiest way to proceed now is to recast the action on the vector space as an exponential representation:

$$\Omega = e^{\mathcal{F}} \quad \mathcal{F} \in adjoint\ representation$$
$$O = e^{F} \quad F \in fundamental\ representation$$

It is always possible to Taylor expand an element of the Lie Algebra around the identity:

$$Tr\Omega = Tre^{\mathcal{F}} = Tr\left(1 + \mathcal{F} + \frac{\mathcal{F}^2}{2!} + \frac{\mathcal{F}^3}{3!} + ..\right)$$

$$= \frac{1}{2}\left\{\left[tr\left(1 + F + \frac{F^2}{2!} + \frac{F^3}{3!} + ..\right)\right]^2 - tr\left(1 + (2F) + \frac{(2F)^2}{2!} + \frac{(2F)^3}{3!} + ..\right)\right\}$$

where the previous identity (2.3.1.4) has been used in the second step. It is now an easy task to retrieve the trace identities, since what is left to do is comparing the terms with same order:

Table 2: *Trace Identities for SO(n)*

$$Tr1 = \frac{1}{2}\left[(tr1)^2 - tr1\right] = \frac{n}{2}(n-1)$$
$$TrF = 0$$
$$TrF^2 = (n-2)trF^2$$
$$TrF^4 = (n-8)trF^4 + 3(trF^2)^2$$
$$TrF^6 = (n-32)trF^6 + 15(trF^2)(trF^4)$$

where we have used $tr1 = n$. For simplicity, we have dropped the symbol $\mathcal{F}$; it is intended, however, that the element in the LHS of which we calculate the trace always lives in the adjoint representation.

The box of trace identities for SO($n$) allows us to answer a major question that was left open a couple of paragraphs ago. The only *SO(n)* group that is allowed is *SO(32)* because of the last identity; moreover the identities used in the introduction are retrieved from the above ones. For $n \neq 32$, indeed, $TrF^6$ does not factorize and therefore remains unchanged in (1.2.3), which consequently cannot be factorized in 8-forms and 4-forms.

If the total anomaly polynomial is not factorized, the anomaly cannot be canceled via the Green-Schwarz mechanism.

*SO(32)* also contains 496 generators and is coupled to a known string theory. All these conditions guarantee a special place in subset B of Fig.1 to this very special gauge group. As for the remaining *SO(n)* groups, they cannot even be included in subset A of "groups where the anomaly is cancelled" for the reasons just cited.

### 2.3.2 Sp($n$)

$Sp(n)$ is a class of groups that shares several properties with $SU(n)$. As far as our topic is concerned, the major and crucial difference between the two gauge groups lies in the carrier space: it is anti-symmetric for $SO(n)$ and symmetric for $Sp(n)$. The analysis of the latter class goes hand in hand with that of the former, with the, at first, harmless-looking difference in the sign of the relation between the traces in adjoint and fundamental representations. For $Sp(n)$ we have:

$$\Omega_{kl}^{mn} = O_k^m O_l^n + O_k^n O_l^m$$

which leads to:

$$Tr\Omega = \frac{1}{2}[(trO)^2 + tr(O^2)]$$

At this point the expression can be Taylor expanded in much the same way as before:

$$Tr\Omega = Tre^{\mathcal{F}} = Tr\left(1 + \mathcal{F} + \frac{\mathcal{F}^2}{2!} + \frac{\mathcal{F}^3}{3!} + ..\right)$$
$$= \frac{1}{2}\left\{\left[tr\left(1 + F + \frac{F^2}{2!} + \frac{F^3}{3!} + ..\right)\right]^2 + tr\left(1 + (2F) + \frac{(2F)^2}{2!} + \frac{(2F)^3}{3!} + ..\right)\right\}$$

So that, finally, the trace relations for $Sp(n)$ can be found.

Table 3:*Trace Identities for Sp(n)*

$$Tr1 = \frac{1}{2}[(tr1)^2 + tr1] = \frac{n}{2}(n+1)$$
$$TrF = 0$$
$$TrF^2 = (n+2)trF^2$$
$$TrF^4 = (n+8)trF^4 + 3(trF^2)^2$$
$$TrF^6 = (n+32)trF^6 + 15(trF^2)(trF^4)$$

And here is the caveat. There can be no gauge group for which $n$ cancels the first term $trF^6$ of the last identity. Therefore, as the term survives, the anomaly cannot be canceled via G.S. mechanism.

### 2.3.3 $SU(n)$

In the case of $SU(n)$ the action on the carrier space is given by [7]:

$$(Uv)_i^j = U_i^{j;k}{}_l v_k^l = u_i^k (u^*)_l^j v_k^l$$

whereas the relation between adjoint and fundamental actions is obtained by taking the direct product of **n** and **n\*** in $SU(n)$ minus the trace:

$$U_i^{j;k}{}_l = u_i^k (u^*)_l^j - \frac{1}{n}\delta_i^j \delta_l^k$$

Upon calculating the traces we obtain:

$$Tr\, U = (tr\, u)(tr\, u^*) - 1 \qquad (2.3.3.1)$$

since $tr\delta_i^j \delta_l^k = n$. The relation (2.3.3.1) is quite what we wanted. Analogously to $SO(n)$ the actions are recast in terms of elements of the gauge algebra.

$$U = e^{\mathcal{F}} \quad \mathcal{F} \in adjoint\ representation$$
$$u = e^{F} \quad F \in fundamental\ representation$$

Expanding LHS and RHS of relation (2.3.3.1) the following is obtained:

$$Tr\left(1 + \mathcal{F} + \frac{\mathcal{F}^2}{2!} + \frac{\mathcal{F}^3}{3!} + ..\right)$$
$$= tr\left(1 + F + \frac{F^2}{2!} + \frac{F^3}{3!} + ..\right) tr\left(1 + F + \frac{F^2}{2!} + \frac{F^3}{3!} + ..\right)^* - 1$$

It is important to note at this point that $F$ is anti-hermitian and therefore satisfies $F^* = -F$. Expanding the RHS in light of this property and then comparing the terms with same power leads to the trace identities for $SU(n)$:

Table 4: *Trace identitites for $SU(n)$*

$$Tr1 = n^2 - 1$$
$$TrF = 0$$
$$TrF^2 = 2n(trF^2)$$
$$TrF^4 = 2n(trF^4) + 6(trF^2)^2$$
$$TrF^6 = 2n(trF^6) + 30(trF^4)(trF^2)$$
$$\qquad -20(trF^3)^2$$

We see that the trace of the sixth order curvature is not cancelled and therefore all $SU(n)$ theories are anomalous.

## 2.4 Anomalies in Exceptional groups

### 2.4.1 *$G_2$*

$G_2$ has been widely studied, therefore many of its properties are well-known. It is known, for example, that the group possesses 14 generators and that the fundamental rep is **7** [7].
For our purposes, we will only be concerned with the maximal subgroup $SU(2)$. The action of the maximal subgroup on the **7**={$x^\alpha, \bar{x}_\alpha, y$}, where $(x^\alpha)^* = \bar{x}_\alpha$ represent hermitian spinors and $y$ represents a scalar, defines transformations closed under a Lie Algebra with generators $T_i$ (see [7] for more details). The decomposition of adjoint and fundamental reps under *$SU(3)$* yields:

$$\mathbf{14} \longrightarrow \mathbf{8} + \mathbf{3} + \mathbf{3}^*, \qquad adjoint$$
$$\mathbf{7} \longrightarrow \mathbf{3} + \mathbf{3}^* + \mathbf{1}, \qquad fundamental$$

$G_2$ has only two Casimir operators, $C_2$ and $C_6$: therefore both of them play a role in determining $TrF^6$. Specifically, the identity of the sixth power must satisfy the following relation:

$$TrF^6 = \alpha\, trF^6 + \beta\, (trF^2)^3 \qquad\qquad (2.4.1.1)$$

Let us have a look at this equation. As repeatedly emphasized, anomaly cancellation can only occur when $TrF^6$ factorizes. In the method reported in this section, such statement is equivalent to saying that $\alpha$ is vanishing: if this is the case, indeed, $TrF^6$ can be expressed in terms of traces of lower power in the fundamental representation. Conversely, if $\alpha \neq 0$ the theory is anomalous.
The generators of the Lie Algebra are the starting point of the discussion for $G_2$; this is again due to their relative ease to use compared to the general elements of the Lie Algebra such as the curvature $F$. Throughout the next paragraphs, the calculations that will be performed using the generators. Indeed, they are easier to use and a non-vanishing trace of the sixth power of the generator in the adjoint representation bears the same consequences of a non-vanishing $\alpha$.
In our case the generators of the maximal subalgebra are:

$$T_1 = -\frac{i}{2}\begin{pmatrix} 0 & 1 & 0 \\ 1 & 0 & 0 \\ 0 & 0 & 0 \end{pmatrix} \qquad T_2 = -\frac{i}{2}\begin{pmatrix} 0 & -i & 0 \\ i & 0 & 0 \\ 0 & 0 & 0 \end{pmatrix}$$

$$T_3 = -\frac{i}{2}\begin{pmatrix} 1 & 0 & 0 \\ 0 & -1 & 0 \\ 0 & 0 & 0 \end{pmatrix} \qquad T_4 = -\frac{i}{2}\begin{pmatrix} 0 & 0 & 1 \\ 0 & 0 & 0 \\ 1 & 0 & 0 \end{pmatrix}$$

$$T_5 = -\frac{i}{2}\begin{pmatrix} 0 & 0 & -i \\ 0 & 0 & 0 \\ i & 0 & 0 \end{pmatrix} \qquad T_6 = -\frac{i}{2}\begin{pmatrix} 0 & 0 & 0 \\ 0 & 0 & 1 \\ 0 & 1 & 0 \end{pmatrix}$$

$$T_7 = -\frac{i}{2}\begin{pmatrix} 0 & 0 & 0 \\ 0 & 0 & -i \\ 0 & i & 0 \end{pmatrix} \quad T_8 = -\frac{i}{2\sqrt{3}}\begin{pmatrix} 1 & 0 & 0 \\ 0 & 1 & 0 \\ 0 & 0 & -2 \end{pmatrix}$$

We can then retrieve an expression for the second, fourth and sixth power of a generator and calculate the relevant traces. The trace of the adjoint is calculated using the identities of $SU(n)$; the choice of course being related to the maximal subalgebra $SU(3)$.

Since relation (2.4.1.1) contains two constants, the procedure described above must be carried for two generators, so to obtain a system of two equations in two variables: the generators of interest for us are $T_3$ and $T_8$. We begin by calculating a first expression in two variables using the former generator:

$$T_3^2 = -\frac{1}{4}\begin{pmatrix} 1 & 0 & 0 \\ 0 & 1 & 0 \\ 0 & 0 & 0 \end{pmatrix} \qquad\qquad T_3^6 = -\frac{1}{64}\begin{pmatrix} 1 & 0 & 0 \\ 0 & 1 & 0 \\ 0 & 0 & 0 \end{pmatrix}$$

and therefore,

$$Tr(T_3^6) = -\frac{1}{64} - \frac{1}{64} = -\frac{2}{64}$$
$$Tr(T_3^{6^*}) = -\frac{1}{64} - \frac{1}{64} = -\frac{2}{64}$$

The last trace to obtain before calculating the total trace of the adjoint is $Tr(T_8^6)$. Upon using the trace identities of $SU(n)$ the following result is obtained:

$$Tr(T_8^6) = -2 - \frac{1}{16}$$

and in conclusion the trace of the adjoint representation is given by:

$$Tr(T_3^6) = -2 - \frac{1}{16} - \frac{2}{64} - \frac{2}{64} = -\frac{17}{8} \qquad (2.4.1.2)$$

As far as the traces of the fundamental representation go, we have:

$$tr(T_3^6) = tr(T_3^6) + tr(T_3^{*6}) = -\frac{4}{64}, \quad tr(T_3^2) = tr(T_3^2) + tr(T_3^{*2}) = -1 \quad (2.4.1.3)$$

Substituting equations (2.4.1.3) and (2.4.1.2) into the relation (2.4.1.1) the first expression up to two constants is obtained:

$$-\frac{17}{8} = \alpha\left(-\frac{1}{16}\right) + \beta(-1) \qquad (2.4.1.4)$$

This is clearly not enough to determine $\alpha$ and $\beta$, but it shows the path to follow to obtain a second expression of such type.

Now let us calculate the square, fourth power and sixth power of $\lambda_8$.

$$T_8^2 = -\frac{1}{12}\begin{pmatrix} 1 & 0 & 0 \\ 0 & 1 & 0 \\ 0 & 0 & 4 \end{pmatrix} \quad T_8^3 = \frac{i}{24\sqrt{3}}\begin{pmatrix} 1 & 0 & 0 \\ 0 & 1 & 0 \\ 0 & 0 & -8 \end{pmatrix}$$

$$T_8^4 = \frac{1}{144}\begin{pmatrix} 1 & 0 & 0 \\ 0 & 1 & 0 \\ 0 & 0 & 16 \end{pmatrix} \quad T_8^6 = -\frac{1}{12^3}\begin{pmatrix} 1 & 0 & 0 \\ 0 & 1 & 0 \\ 0 & 0 & 64 \end{pmatrix}$$

Hence, the traces of the fundamental representation are given by

$$tr(T_8^2) = -1, \qquad tr(T_8^6) = \left(-\frac{1}{12}\right)^3(66+66) = -\frac{11}{144}$$

whereas the trace of the adjoint is:

$$Tr(T_8^6) = -\frac{27}{16} - \frac{11}{144} = -\frac{127}{72}$$

The expression obtained in this case is:

$$-\frac{127}{72} = \alpha\left(-\frac{11}{144}\right) + \beta(-1) \qquad (2.4.1.5)$$

It is clear that this is incompatible with (2.4.1.4). A first, naïve, thought would be to consider the only case of interest for us, namely $\alpha = 0$. Such particular result leads to two different values of $\beta$ in the system; clearly a contradiction. To be more precise, the system of equations yields $\alpha = -26$ and $\beta = \frac{15}{4}$. Relation (2.4.1.1) then becomes:

$$TrF^6 = -26\, trF^6 + \frac{15}{4}\,(trF^2)^3 \qquad (2.4.1.6)$$

and does not factorize. In conclusion, $G_2$ is not anomalous.

Incidentally, one might wonder if the trace $trF^6$ of (2.4.1.1) further factorizes into $(trF^2)^3$; in other words, one might wonder if $trF^6$ is a primitive Casimir. The above discussion rules out this possibility since $\alpha$ and $\beta$ are determined *uniquely*. Suppose, conversely, that it is possible to factorize $trF^6$ as follows:

$$trF^6 = \kappa(trF^2)^3 \qquad (2.4.1.7)$$

Then inserting (2.4.1.7) into (2.4.1.6) we obtain:

$$trF^6 = \left(-26\kappa + \frac{15}{4}\right)(trF^2)^3$$

So that in this case the most general (2.4.1.1) has solutions:

$$(\alpha, \beta) = \left(0, -26\kappa + \frac{15}{4}\right)$$

These values are clearly differing from the unique constants obtained with the general method, so that the initial assumption is not valid and $trF^6$ is a primitive Casimir by contradiction.

This proof holds whenever the constants of $TrF^6$ are determined uniquely.

As we shall see this is the case for all the exceptional groups, so that $trF^6$ is a primitive Casimir in all the groups that we will be interested in.

### 2.4.2 $F_4$

$F_4$ has 52 generators and $SO(9)$ as the maximal subalgebra. The fundamental representation is **26**. The decomposition under $SO(9)$ is:

$$\begin{aligned} \mathbf{52} &\rightarrow \mathbf{36 + 16}, && adjoint \\ \mathbf{26} &\rightarrow \mathbf{9 + 16 + 1}, && fundamental \end{aligned}$$

$F_4$ has also rank 4, therefore it possesses 4 Casimir operators, which are: $C_2, C_6, C_8, C_{12}$. [8] In light of this, the expression for the trace is still (2.4.1.1). The problem is still to check if $\alpha \neq 0$. The two representations of the generators chosen are:

$$T_1 = \begin{pmatrix} 0 & 1 & 0 & \\ -1 & 0 & 0 & \cdots \\ 0 & 0 & 0 & \\ & \vdots & & \ddots \end{pmatrix}_{9\times 9} \qquad\qquad T_2 = \begin{pmatrix} 0 & 1 & 1 & \\ -1 & 0 & 0 & \cdots \\ -1 & 0 & 0 & \\ & \vdots & & \ddots \end{pmatrix}_{9\times 9}$$

The spinor representation is given by $T_{sp}{}^2 = -\frac{1}{4}I_{16\times 16}$ [7].

The trace of the fundamental representation is the sum of the contribution from the generator and that of the spinor representation. Therefore we have:

$$tr(T_1)^2 = -2 - \frac{16}{4} = -6$$

$$tr(T_1)^6 = -2 - \frac{16}{64} = -\frac{9}{4}$$

and following the trace identities for $SO(n)$ we obtain as the adjoint trace:

$$Tr(T_1)^6 = -14 - \frac{16}{64} = -\frac{57}{4}$$

which leads to an expression for the trace in two constants:

$$\frac{57}{4} = \alpha \left(\frac{9}{4}\right) + \beta (216)$$

The same procedure is repeated for the second generator. The spinor representation is in this case $T_{sp}{}^2 = -\frac{1}{2} I_{16 \times 16}$. One then obtains:

$$tr(T_2)^2 = -4 - 8 = -12 \qquad (2.4.2.1)$$

$$tr(T_2)^6 = -10 - 2 = -12$$

$$Tr(T_2)^6 = -132 \qquad (2.4.2.2)$$

Where the last trace has been found with the usual relation for the trace identities of $SO(n)$. The second trace identity up to two constants is then:

$$132 = \alpha(12) + \beta(1728)$$

and it is clear that $\alpha$ is non-vanishing. The system of equations results in $\alpha = -3, \beta = \frac{7}{72}$, so that the trace identity (2.4.1.1) becomes:

$$TrF^6 = -3\ trF^6 + \frac{7}{72}\ (trF^2)^3$$

Therefore $F_4$ is anomalous.

### 2.4.3 $E_6$

This group has adjoint representation **78** and fundamental **27**. There are several subgroups that one could use to decompose such representations.
Two examples are $SO(10) \times U(1)$ and $Usp(8)$, under which the decomposition can be found in [9]. Another suitable subgroup is $F_4$ and we will use it since most of the calculations involved have already been carried out in section 2.4.2. Under $F_4$ the decomposition is:

$$\begin{aligned} \mathbf{78} &\rightarrow \mathbf{26} + \mathbf{52}, && adjoint \\ \mathbf{27} &\rightarrow \mathbf{1} + \mathbf{26}, && fundamental \end{aligned}$$

As far as the Casimir operators are concerned, the relevant ones are $C_2$ and $C_6$. Hence, the trace to find is still (2.4.1.1). The trace of the fundamental rep for generator 1 is the same as (2.4.2.1), since the factor 1 does not contribute to the trace. We see also that the trace of the adjoint is the sum of traces (2.4.2.1) and (2.4.2.2), already calculated previously. The first relation then is easily obtained:

$$144 = \alpha(12) + \beta(1728)$$

Similary, the traces fundamental and adjoint representation are calculated using $tr(T_1)^6$ and $Tr(T_2)^6$ from the previous section to obtain:

$$\frac{33}{2} = \alpha \left(\frac{9}{4}\right) + \beta \left(\frac{9}{4}\right)^3$$

The system of equations gives $\alpha = \frac{5308}{741}, \beta = \frac{224}{6669}$. Anomaly cancellation, even in this case, is not possible. The trace identity becomes:

$$TrF^6 = \frac{5308}{741} \ trF^6 + \frac{224}{6669} \ (trF^2)^3$$

### 2.4.4 $E_7$

The next structure encountered in our journey through the exceptional groups is $E_7$. The group has 133 generators and $SU(8)$ as a maximal subgroup. It is also important to know that the fundamental representation is **56**. Under $SU(8)$ the decomposition is as follows:

$$\mathbf{133} \rightarrow \mathbf{63} + \mathbf{70}, \qquad adjoint$$
$$\mathbf{56} \rightarrow \mathbf{28} + \mathbf{28}^*, \qquad fundamental$$

The relevant Casimir are unchanged, hence there is no need to modify the structure of $TrF^6$. Decomposing under the maximal subgroup and taking into account the dimensionality of the representations we obtain:

$$Tr_{\mathbf{63}}F^6 + tr_{\mathbf{70}}F^6 = \alpha \ (2tr_{\mathbf{28}}F^6) + \beta (2tr_{\mathbf{28}}F^2)^3$$

where we have exploited the fact that the fundamental representation is real. The method used for this exceptional group differs slightly from the previous ones. To check that $\alpha \neq 0$ we calculate the coefficients of the sixth power element of the Lie Algebra for both first and second term in the LHS. Assuming $\alpha = 0$ such coefficient should vanish: checking that this is or is not the case is equivalent to checking a non-vanishing $\alpha$ and therefore the factorization or non-factorization of the anomaly polynomial.

As generators we choose any linear combination $T$ of the seven independent generators of $SU(8)$, labeled $t_i$ for $i$=1,2,...8 [7]. $T$ is therefore an element of the Lie Algebra and we can use the last trace identity in table 4 to obtain:

$$Tr_{\mathbf{63}}T^6 = 8 \cdot 2\sum_i t_i^6 + 15 \sum_i t_i^2 \sum_i t_i^4 + 15 \sum_i t_i^4 \sum_i t_i^2 - 20 \left(\sum_i t_i^3\right)^2$$

$$= 16\sum_i t_i^6 + 30 \sum_i t_i^2 \sum_i t_i^4 - 20 \left(\sum_i t_i^3\right)^2 \qquad (2.4.4.1)$$

A similar argument is used for $tr_{\mathbf{70}}F^6$. In this case, however, the states are labeled $t_i, t_j, t_k, t_l$ with $i < j < k < l$. Let us focus on the sixth power term. The restricted sum of the states can be recast as a combination of unrestricted sums [7].

$$\sum_{i<j<k<l} \left(t_i + \ t_j + t_k + t_l\right)^6 =$$

$$\frac{1}{24}\Big(\sum_{i,j,k,l}(t_i + t_j + t_k + t_l)^6 - 6\sum_{i=j,k,l}(2t_i + t_k + t_l)^6 +$$

$$3\sum_{i=j,k=l}(2t_i + 2t_k)^6 + 8\sum_{i=j=k,l}(3t_i + t_l)^6 - 6\sum_{i=j=k=l}(4t_i)^6\Big)$$

$$(2.4.4.2)$$

Using (2.4.4.1) and (2.4.4.2) we can finally collect the two coefficients of the sixth power terms.

$$Tr_{\mathbf{63}}T^6 \propto c_1\sum_i t_i^6, \quad tr_{\mathbf{70}}T^6 \propto c_2\sum_i t_i^6$$

$c_1$ is readily obtained by inspection of (2.4.4.1). Clearly, $c_1 = 16$. A more involved calculation is needed for $c_2$.

$$c_2 = \frac{1}{24}[1^6 + 1^6 + 1^6 + 1^6 - 6(2^6 + 1^6 + 1^6) + 3(2^6 + 2^6)$$

$$+8(3^6 + 1^6) - 6(4^6) = 3^5 - 2^{10}$$

The appropriate case for us is $\alpha = 0$. Assuming this is the case and considering only the sixth powers, the RHS of relation (2.4.1.1) vanishes: the first term is indeed zero by assumption, the second is not proportional to $t_i^6$.
The condition then becomes:

$$(c_1 + c_2)\sum_i t_i^6 = 0 \cdot \sum_i t_i^6 = 0$$

But clearly $c_1 + c_2 \neq 0$ as it has been calculated just above. We conclude that $\alpha$ must be different than zero and that consequently the factorization of $TrF^6$ is not possible.

### 2.4.5 $E_8$

We end our discussion on exceptional groups with the one relevant to anomaly-free gauge theories. In section 1.2 of the Introduction it was discussed how $E_8 \times E_8$ cancels the anomaly via Green-Schwarz mechanism. It was also anticipated that $U(1)^{496}$ and $U(1)^{248} \times E_8$ cancel the anomalies. It is now time to spare a thought on the last statement. Why does $E_8$ behave differently from its exceptional siblings?
The answer lies in the Casimir operators of $E_8$. These are $C_i$ for $i$=2,8,12,14,18,20,24,30. Therefore the group does not possess $C_6$. Consequently, the factorization of $TrF^6$ *is* possible because there *is not* such factor at all. In the literature [2] one can see that $TrF^4$ and $TrF^2$ are non-vanishing and factorizable [see the trace identities in (1.3.1)].
For the above reasons $E_8$ provides a good base to develop anomaly-free models. In the introduction $E_8 \times E_8$ was considered, however the reader should also remember that the group $U(1)$ behaves differently from its class $U(n)$. Indeed, this gauge group cancels the anomaly (a good reference is chapter 12 of [5]). As a corollary any 496-dimensional theory constructed solely upon multiplication of $U(1)$ is anomaly-free. Clearly, the only possible way is to consider the anomaly-free $U(1)^{496}$ theory.

Finally, we could combine $E_8$ and $U(1)$ in a 496-dimensional theory that is anomaly-free for the reasoning just cited. Again, there is only possible way to do this, e.g. $U(1)^{248} \times E_8$. For an algebraic treatment of these theories, e.g. factorization via the Green-Schwarz mechanism, consult the appendix of this paper. There, the Green Schwarz terms are retrieved for both the groups in the swampland, in a manner that is similar to the analysis carried out for $E_8 \times E_8$ and $SO(32)$ in the introduction.

## 2.5 Some "strange" low-dimensional groups

It seems that the method used above provides a way to answer positively to the initial question: are the groups where the anomaly is cancelled only four? However, care must be taken in analyzing the method.

Indeed, we start our discussion assuming that $TrF^6$ factorizes if the first term $\alpha \, trF^6$ vanishes. This is checked demonstrating that $\alpha \neq 0$ for any class of groups. However, a vanishing $\alpha$ does not logically preclude the idea that $trF^6$ itself might factorize in lower order powers of $F$. Such factorization is a property of some low-dimensional groups that, nicely, can all be traced to some $SU(n)$ for $n \leq 5$, $Sp(2)$ and products thereof.

The method presented previously, therefore, does not stand for them as formulated above and anomaly cancellation must be checked differently, in such a way to account for the factorization of $TrF^6$.

Fortunately enough, the Green-Schwarz mechanism holds for any group and thus also for low dimensional $SU(n)$ groups and $Sp(2)$. It seems a good idea to start from this reassuring mechanism and we will do it employing it for $Sp(2)$ first.

### 2.5.1 $Sp(2)$

The goal here is to factorize the total anomaly polynomial (1.2.3), in line with the general idea of the mechanism explained in section 1.3. The three terms of the polynomial that factorize in lower order ones are $TrF^6$, $TrF^4$ and $TrF^2$.

Specifically, the structure of the factorization is such that:

$$TrF^2 = C_1 \cdot tr \, F^2$$

$$TrF^4 = C_2 \cdot (tr \, F^2)^2 + C_3 \cdot tr \, F^4$$

$$TrF^6 = C_4 \cdot (tr \, F^2)^3 + C_5 \cdot tr \, F^4 tr \, F^2$$

The generators $T$ used for checking $Sp(2)$ are:

$$T_1 = \begin{pmatrix} 0 & i \\ i & 0 \end{pmatrix} \qquad T_2 = \begin{pmatrix} i\sigma_1 & 0 \\ 0 & i\sigma_3 \end{pmatrix}$$

Only one generator is needed for $TrF^2$, as it contains one constant, and the factorization is trivially found using the results in Table 3:

$$TrF^2 = 6\, tr\, F^2 \qquad (2.5.1.1)$$

$TrF^4$ needs two sets of calculations performed on the two generators. Again, the LHS is calculated using the general trace identity of the fourth power for $Sp(n)$ with $n = 2$. The two generators yield a system of equations with variables $C_2$ and $C_3$ that, once solved, allows us to write down the trace identity for $TrF^4$:

$$TrF^4 = 3(tr\, F^2)^2 + 12\, tr\, F^4 \qquad (2.5.1.2)$$

A similar argument is brought upon for the trace of the sixth power, which reads:

$$TrF^6 = -\frac{9}{2}(tr\, F^2)^3 + 42\, tr\, F^4 tr\, F^2 \qquad (2.5.1.3)$$

For simplicity we now set $n = 496$ in (1.2.3), so to obtain the total anomaly polynomial of interest for the factorization (we adapt to the notation of [2] here):

$$I_{12}|_{n=496} = -\frac{1}{15}TrF^6 + \frac{1}{24}TrF^4 trR^2 - \frac{1}{960}TrF^2[4trR^4 + 5(trR^2)^2]$$
$$+ \frac{1}{8}trR^2 trR^4 + \frac{1}{32}(trR^2)^3 \qquad (2.5.1.4)$$

Relations (2.5.1.1), ( 2.5.1.2) and ( 2.5.1.3) are now plugged into (2.5.1.4 ):

$$I_{12}{}^{Sp(2)} = \frac{9}{30}(tr\, F^2)^3 - \frac{42}{15}trF^2 trF^4 + \frac{1}{8}(tr\, F^2)^2 trR^2 + \frac{1}{2}trR^2 trF^4$$
$$- \frac{1}{40}trF^2 trR^4 - \frac{1}{32}trF^2(trR^2)^2 + \frac{1}{8}trR^2 trR^4 + \frac{1}{32}(trR^2)^3$$

$$(2.5.1.5)$$

The question is now whether or the polynomial above factorizes.

Before going any further, it is compelling to report that the Green-Schwarz in its original sense is the application of the counterterm $\Delta\Gamma$ in heterotic supergravity: this theory has a 12-degree Chern class $I_{12}$ that factorizes as $I_8 \wedge I_4$; that is precisely how the physical electric and magnetic currents respectively factorize. The gauge theories that we have been considering in this paper fall in this category, and therefore when we check factorization we also have to check that the following constraint is satisfied:

$$I_{12} = Y_4 \wedge X_8 \qquad (2.5.1.6)$$

Of course, nothing prevents the polynomial (2.5.1.4) from factorizing without condition (2.5.1.6). However, the resulting gauge theories would not be consistent from the string viewpoint. In light of this, we will be concerned only with theories that factorize and satisfy the constraint.

Let us turn our attention to $Sp(2)$ again. $F$ and $R$ have by definition degree 2. In principle, therefore, the 4-degree and 8-degree terms can be separated up to seven constants:

$$Y_4 \wedge X_8 =$$
$$(c_1 trF^2 + c_2 trR^2) \cdot [c_3(trF^2)^2 + c_4(trR^2)^2 + c_5 trR^2 trF^2 + c_6 trF^4 + c_7 trR^4]$$

$$(2.5.1.7)$$

What is not sure, however, is whether or not seven constants that satisfy the equation simultaneously exist. This can be checked by inspection setting one of the variables to be $c_1 = \frac{1}{10}$. By consequent expansion and comparison of terms in equations (2.5.1.7) and (2.5.1.5) we obtain: $c_3 = 3$, $c_2 = \frac{1}{24}$, $c_7 = \frac{1}{192}$, $c_6 = \frac{1}{48}$ and $c_4 = \frac{1}{768}$, which have been computed comparing terms $(tr\, F^2)^3$, $(tr\, F^2)^2 trR^2$, $trR^2 trR^4$ and $(trR^2)^3$ respectively.

Let us have a look at what we have so far:

$$Y_4 \wedge X_8 = \left(\frac{1}{10}\, trF^2 + \frac{1}{24}\, trR^2\right)$$
$$\cdot \left[3(trF^2)^2 + \frac{1}{768}(trR^2)^2 + c_5 trR^2 trF^2 + \frac{1}{48}\, trF^4 + \frac{1}{192}\, trR^4\right]$$

The only remaining constant is $c_5$. However, we can see that it can be retrieved comparing two addends, $(tr\, F^2)^2 trR^2$ and $trF^2(trR^2)^2$. In fact, the system satisfying $c_5$ is overdetermined as there are two equations for one variable. If these are in accordance a unique solution for the constant can be found and thus the total anomaly polynomial can be factorized completing the work done so far. However, comparing the $(tr\, F^2)^2 trR^2$ terms the result reads $c_5 = 0$.

A contradiction is obtained when this is inserted in the equation comparing the latter term $trF^2(trR^2)^2$, thus proving that the anomaly of $Sp(2)$ cannot be cancelled through the G.S. mechanism.

### 2.5.2 $SU(2)$

In section 2.5.1 we have developed a method to construct a factorized polynomial up to six constants and to check it for a particular theory. Of course the particular theory need not be $Sp(2)$ and we herein show that the argument stands for the next group considered, $SU(2)$.

The structures of the factorization of higher order traces is even simpler than $Sp(2)$ as one single constant is present in each relation:

$$TrF^2 = \zeta_1 \cdot tr\, F^2$$

$$TrF^4 = \zeta_2 \cdot (tr\, F^2)^2$$

$$TrF^6 = \zeta_4 \cdot (tr\, F^2)^3$$

The generator is $T = i\sigma = \begin{pmatrix} i & 0 \\ 0 & -i \end{pmatrix}$ and the trace identities are those of table 4, with the particular choice $n = 2$:

$$TrF^2 = 4tr\, F^2 \qquad (2.5.2.1)$$

$$TrF^4 = 8(tr\, F^2)^2 \qquad (2.5.2.2)$$

$$TrF^6 = 16(tr\, F^2)^3 \qquad (2.5.2.3)$$

The total anomaly polynomial for such a theory is therefore:

$$I_{12}{}^{SU(2)} = -\frac{16}{15}(tr\, F^2)^3 + \frac{1}{3}(tr\, F^2)^2 trR^2 - \frac{1}{240}trF^2[4trR^4 + 5(trR^2)^2]$$
$$+ \frac{1}{8}trR^2 trR^4 + \frac{1}{32}(trR^2)^3$$

and again we look for a factorization satisfying (2.5.1.6). This has the form:

$$(c_1 trF^2 + c_2 trR^2) \cdot [c_3(trF^2)^2 + c_4(trR^2)^2 + c_5 trR^2 trF^2 + c_6 trR^4]$$

Let us begin by setting $c_2 = \frac{1}{2}$. By comparing terms with a factor $trR^2 trR^4$ we readily obtain $c_6 = \frac{1}{4}$. Then, in much the same way as described above, the two constants are used as bases to obtain: $c_1 = -\frac{1}{15}$, $c_3 = 16$ and $c_5 = \frac{1}{16}$. Now the situation looks like this:

$$\left(-\frac{1}{15}trF^2 + \frac{1}{2}\,trR^2\right) \cdot \left[16(trF^2)^2 + c_4(trR^2)^2 + \frac{1}{16}\,trR^2 trF^2 + \frac{1}{4}\,trR^4\right]$$

$c_4$ is determined by an usual overdetermined system of equations obtain by comparison of the coefficients of $trF^2(trR^2)^2$ and $trR^2(trF^2)^2$. Unfortunately, even this time the system leads to a contradiction and $SU(2)$ cannot therefore be factorized and cancelled via G.S. mechanism.

### 2.5.3 $SU(3)$

This is a critical group. The factorization of higher order traces is as follows:

$$TrF^2 = X_1 \cdot tr\, F^2$$

$$TrF^4 = X_2 \cdot (tr\, F^2)^2$$

$$TrF^6 = X_3 \cdot (tr\, F^2)^3 + X_4 \cdot (tr\, F^3)^2$$

Upon using $T = \begin{pmatrix} a & 0 & 0 \\ 0 & b & 0 \\ 0 & 0 & a-b \end{pmatrix}$ as generator and the trace identities of table 4 with $n = 3$ we get:

$$TrF^2 = 6\ tr\ F^2$$

$$TrF^4 = 9\ (tr\ F^2)^2$$

$$TrF^6 = \frac{33}{2}(tr\ F^2)^3 + 22\ (tr\ F^3)^2$$

Inserting such relations into the anomaly polynomial we obtain:

$$I_{12}{}^{SU(3)} = -\frac{11}{10}(tr\ F^2)^3 + \frac{22}{15}\ (tr\ F^3)^2 + \frac{3}{8}(tr\ F^2)^2 trR^2 - \frac{1}{40}trF^2 trR^4$$
$$-\frac{1}{32}trF^2(trR^2)^2 + \frac{1}{8}trR^2 trR^4 + \frac{1}{32}(trR^2)^3$$

$$(2.5.3.1)$$

And already it can be seen that the second term is proportional to $tr\ F^3$ having degree 6. Factorization of this polynomial might be possible, but it is impossible for it to satisfy condition (2.5.1.6). Therefore a non-vanishing $trF^3$ term leads to the non-factorizability of (2.5.3.1). The theory is anomalous.

### 2.5.4 $SU(4)$ and $SU(5)$

The trace identity of the sixth power curvature is for both theories:

$$TrF^6 = \alpha \cdot (tr\ F^2)^3 + \beta \cdot (tr\ F^3)^2 + \gamma \cdot tr\ F^4 tr\ F^2 \qquad (2.5.4.1)$$

An argument similar to that used for $SU(3)$ can be exploited. It is now clear that the problem now reduces to the preliminary question: does the coefficient of $trF^3$ vanish? If $\alpha \neq 0$, indeed, we cannot even try to cancel the anomaly via the G.S. mechanism as we are prevented from doing it by condition (2.5.1.6).
The way this is formulated reminds us of the method used to check classical and exceptional groups in sections 2.3 and 2.4, where the question was: does the coefficient of $trF^6$ vanish?
To proceed, then, we will first the question answer whether $\alpha = 0$ or not; consequently, we demonstrate that the G.S. mechanism cannot be used.

### 2.5.4.1 $SU(4)$

Apart from the above constructed trace of the sixth order (2.5.4.1), $SU(4)$ has the following trace structures:

$$TrF^2 = Y_1\ tr\ F^2$$

$$TrF^4 = Y_2\ (tr\ F^2)^2$$

To obtain $TrF^2$ and $TrF^4$ one generator is sufficient, however $TrF^6$ has 3 constants and thus requires 3 generators. We use the general traceless matrix $T = \begin{pmatrix} a & 0 & 0 \\ 0 & b & 0 \\ 0 & 0 & a-b \end{pmatrix}$ and set three different combinations of $a,b,c$.

$$T_1 = \begin{pmatrix} 1 & 0 & 0 & 0 \\ 0 & 1 & 0 & 0 \\ 0 & 0 & 0 & 0 \\ 0 & 0 & 0 & -2 \end{pmatrix} \qquad T_2 = \begin{pmatrix} 2 & 0 & 0 & 0 \\ 0 & 1 & 0 & 0 \\ 0 & 0 & 0 & 0 \\ 0 & 0 & 0 & -3 \end{pmatrix} \qquad T_3 = \begin{pmatrix} 1 & 0 & 0 & 0 \\ 0 & 1 & 0 & 0 \\ 0 & 0 & 1 & 0 \\ 0 & 0 & 0 & -3 \end{pmatrix}$$

$$(2.5.4.1.1)$$

The trace identities for $SU(n)$ with $n = 4$ are used. The method is always to create a system in 3 equations using the generators and solve it. The results are directly shown below:

$$Tr F^2 = 8 \, tr \, F^2 \qquad\qquad (2.5.4.1.2)$$

$$Tr F^4 = 10 \, (tr \, F^2)^2 \qquad\qquad (2.5.4.1.3)$$

$$Tr F^6 = \frac{1844}{75}(tr \, F^2)^3 + \frac{3}{5}(tr \, F^3)^2 + \frac{804}{25} tr \, F^4 tr \, F^2$$

By inspection $\alpha \neq 0$ and therefore the total anomaly polynomial contains a 6-degree term that does not satisfy (2.5.1.6). The theory is anomalous.

### 2.5.4.2 $SU(5)$

Once the calculations for $SU(4)$ are performed the ones for $SU(5)$ are trivial and the group is easily demonstrated to be anomalous. In fact, the generators of $SU(5)$ are the same as those of $SU(4)$. This means that the generators used before (2.5.4.1.1) can be used in this case as well.

The only condition to be careful about is that the trace identities of $SU(n)$ must now be used with $n$ set to 5.

This of course only affects the trace of the sixth power, so that (2.5.4.1.2) and (2.5.4.1.3) still hold in this case. After the system of equations is computed we have:

$$Tr F^6 = \frac{386}{15}(tr \, F^2)^3 + \frac{3}{4}(tr \, F^3)^2 + \frac{327}{10} tr \, F^4 tr \, F^2$$

and the theory is anomalous for the same reasons explained above.

## 3. Combinations of anomalous theories

### 3.1 Outline of the chapter

In section 2.5.4 we accomplished much in our study of 496 dimensional groups. It is now known that even the low dimensional groups are individually anomalous. However, the G.S. mechanism is a global property of the theory: that is, different contributions can combine in such a way that the total anomaly of the combination factorizes. This is what is done when $U(1)$ theories are combined with $E_8$ or when $E_8$ is combined with itself.

Here we show that if any of the contributions is anomalous, the information of the anomaly is transmitted to the whole group, thus making it anomalous. In doing so, we can conclude that $E_8 \times E_8$ and $E_8 \times U(1)^{248}$ can cancel the anomaly only because the bases theories are anomaly-free.

To analyze this statement, combinations of classical and exceptional groups, $SU(n)$ for $n = 2,3,4,5$ and $Sp(2)$ must be checked. In general, if two theories combine the total trace in the adjoint representation is such that:

$$TrF^6{}_{TOT} = TrF^6{}_{THEORY\ 1} + TrF^6{}_{THEORY\ 2}$$

This helps us rule out combinations involving $SU(n)$ for $n = 3,4,5$ as well as anomalous classical and exceptional groups from the very beginning: in the case of $SU(3), SU(4)$ and $SU(5)$ the 6-degree term $tr\ F^3$ survives for whatever the choice of the latter theory is. In the case of anomalous exceptional and classical groups, the 12-degree term $tr\ F^6$ is the contribution preventing the G.S. mechanism.

Therefore, to investigate how the information of the anomaly is transmitted, anomaly-free theories constructed out of $SU(2)$ and $Sp(2)$ must be studied: generally speaking, it is possible to construct three types of gauge group using such two bases and the anomaly-free $U(1)$. In full generality, the three groups are:

$$G_1 = \prod_{i=1}^{n} SU(2)_i \times U(1)^{496-3n}$$

$$G_2 = \prod_{i=1}^{m} Sp(2)_i \times U(1)^{496-10m}$$

$$G_3 = \prod_{i=1}^{n} SU(2)_i \times \prod_{i=1}^{m} Sp(2)_i \times U(1)^{496-10m-3n}$$

Here we have used the fact that $SU(2)$ and $Sp(2)$ have 3 and 10 dimensions, respectively.

We will show by induction that these groups possess a non-factorizable total anomaly polynomial. In fact, we will start from the simplest case for each of the three groups, e.g. we will start by a combination of only two groups and $U(1)^n$ for some suitable $n$.

By induction it will be clear that adding more groups will not cancel the anomalous contributions and that combinations involving the above groups and $E_8$ are anomalous. This proof will work in a way analogous to falling dominoes: to make all the dominoes fall, one needs a starting piece and a little thrust; in the same way, to rule out all the combinations of anomalous groups one needs a starting anomalous theory, e.g. the ones found in the previous chapter, and a little thrust, e.g. a check that even the simplest combination will not cancel the pre-existing anomaly.

To construct the total anomaly polynomial we will be as general as possible: specifically, we will not start from the anomaly polynomial for $n=496$ that has been used to find those of the low dimensional groups [see (2.5.1.4)], but rather we will be interested in developing a total anomaly polynomial starting from the matter content of each theory.

We also check in the Appendix that starting from the matter content allows recovering the factorization of groups that are known to factorize, hence allowing us to construct Green Schwarz terms for them.

### 3.2 Combinations of $SU(2)$ and $U(1)$

As anticipated, we consider the simplest case $SU(2) \times SU(2) \times U(1)^{490}$. The individual contributions to the total anomaly polynomial are the anomalies of supergravity and of the Yang-Mills theory. To remind the reader, the anomaly arising from supergravity is:

$$I_{12}^{SUGRA} = I_{12}^{gravitino} - I_{12}^{spin\frac{1}{2}}|_{\mathcal{R}=1}$$

where the individual contributions are the ones used already in section 1.2:

$$I_{12}^{gravitino} = \frac{1}{64(2\pi)^5}\left(-\frac{11}{126}trR^6 + \frac{5}{96}trR^4trR^2 - \frac{7}{1152}(trR^2)^3\right) \quad (3.2.1)$$

$$I_{12}^{spin\frac{1}{2}} = \frac{1}{64(2\pi)^5}(tr_{\mathcal{R}}1)\left(\frac{1}{5670}trR^6 + \frac{1}{4320}trR^4trR^2 + \frac{1}{10368}(trR^2)^3\right)$$
$$- \frac{1}{32(2\pi)^5}(tr_{\mathcal{R}}F^2)\left(\frac{1}{360}trR^4 + \frac{1}{288}(trR^2)^2\right) + \frac{1}{1152(2\pi)^5}(tr_{\mathcal{R}}F^4)trR^2$$
$$- \frac{1}{720(2\pi)^5}(tr_{\mathcal{R}}F^6) \quad (3.2.2)$$

As far as the Yang-Mills theory is concerned, we need to calculate an anomaly polynomial for $SU(2)$ and one for $U(1)^{490}$.

$I_{12}^{spin\frac{1}{2}}$ provides a general basis for both theories: in the case of $U(1)^{490}$, the anomaly of the spin is calculated setting $tr_{\mathcal{R}}1 = dimU(1)^{490} \equiv n_1$ and $F = 0$.

In the case of $SU(2)$ the trace relations (2.5.2.1), (2.5.2.2) and (2.5.2.3) are inserted in (3.2.2), with $tr_{\mathcal{R}}1 = dimSU(2) \equiv n_2$. The results are:

$$I_{12}^{U(1)^{490}} = \frac{1}{64(2\pi)^5}\left(\frac{n_1}{5670}trR^6 + \frac{n_1}{4320}trR^4trR^2 + \frac{n_1}{10368}(trR^2)^3\right) \quad (3.2.3)$$

$$I_{12}^{SU(2)} = \frac{1}{64(2\pi)^5}\left(\frac{n_2}{5670}trR^6 + \frac{n_2}{4320}trR^4trR^2 + \frac{n_2}{10368}(trR^2)^3\right)$$
$$- \frac{1}{8(2\pi)^5}(trF^2)\left(\frac{1}{360}trR^4 + \frac{1}{288}(trR^2)^2\right) + \frac{1}{144(2\pi)^5}(trF^2)^2trR^2$$

$$-\frac{1}{45(2\pi)^5}(trF^2)^3 \tag{3.2.4}$$

Starting from (3.2.4) it is trivial to obtain an expression for the anomaly of $SU(2) \times SU(2)$. In fact, the only relevant consideration is to distinguish a factor $R$ common to both theories and a factor $F$ that is different and labeled accordingly.

$$
\begin{aligned}
I_{12}^{SU(2)\times SU(2)} = {} & \frac{1}{64(2\pi)^5}\Big(\frac{n_2+n_3}{5670}trR^6 + \frac{n_2+n_3}{4320}trR^4 trR^2 + \frac{n_2+n_3}{10368}(trR^2)^3\Big) \\
& -\frac{1}{8(2\pi)^5}\big(trF_1{}^2 + trF_2{}^2\big)\Big(\frac{1}{360}trR^4 + \frac{1}{288}(trR^2)^2\Big) \\
& +\frac{1}{144(2\pi)^5}trR^2\Big(\big(trF_1{}^2\big)^2 + (trF_2^2)^2\Big) \\
& -\frac{1}{45(2\pi)^5}\big((trF_1{}^2)^3 + (trF_2{}^2)^3\,\big)
\end{aligned}
\tag{3.2.5}
$$

where $n_3$ denotes the dimensions of the additional $SU(2)$ theory.

The total anomaly of the Yang-Mills theory is simply the sum of contributions to the gauge theory, namely (3.2.3), (3.2.4) and (3.2.5). We use the fact that $n_1 + n_2 + n_3 = 496$ to write:

$$
\begin{aligned}
I_{12}^{YM} = {} & \frac{1}{64(2\pi)^5}\Big(\frac{496}{5670}trR^6 + \frac{496}{4320}trR^4 trR^2 + \frac{496}{10368}(trR^2)^3\Big) \\
& -\frac{1}{8(2\pi)^5}\big(trF_1{}^2 + trF_2{}^2\big)\Big(\frac{1}{360}trR^4 + \frac{1}{288}(trR^2)^2\Big) \\
& +\frac{1}{144(2\pi)^5}\Big(\big(trF_1{}^2\big)^2 + (trF^2{}_2)^2\Big)trR^2 \\
& -\frac{1}{45(2\pi)^5}\big((trF_1{}^2)^3 + (trF_2{}^2)^3\,\big)
\end{aligned}
\tag{3.2.6}
$$

In conclusion, the Yang-Mills is coupled to the supergravity in order to obtain the total anomaly polynomial of $SU(2) \times SU(2) \times U(1)^{490}$.

This, of course, means that the contributions to the total anomaly are $I_{12}^{YM}$, $I_{12}^{gravitino}$ and $-I_{12}^{spin\frac{1}{2}}|_{\mathcal{R}=1}$.

$$
\begin{aligned}
I_{12}^{TOT} = {} & \frac{1}{64(2\pi)^5}\Big(\frac{1}{6}trR^4 trR^2 + \frac{1}{24}(trR^2)^3\Big) \\
& -\frac{1}{8(2\pi)^5}\big(trF_1{}^2 + trF_2{}^2\big)\Big(\frac{1}{360}trR^4 + \frac{1}{288}(trR^2)^2\Big) \\
& +\frac{1}{144(2\pi)^5}\Big(\big(trF_1{}^2\big)^2 + (trF^2{}_2)^2\Big)trR^2 \\
& -\frac{1}{45(2\pi)^5}\big((trF_1{}^2)^3 + (trF_2{}^2)^3\,\big)
\end{aligned}
\tag{3.2.7}
$$

Once the total anomaly has been obtained, a final check of its factorization must be carried out: as usual, the factorization must satisfy $I_{12} = Y_4 \wedge X_8$.
In principle, then, the factorization has the following form:

$$\left(c_1 trR^2 + c_2\ trF_1^2 + c_3\ trF_2^2\right) \cdot [c_4(trR^2)^2 + c_5\left(trF_1^2\right)^2 + c_6\left(trF_2^2\right)^2$$

$$+ c_7 trR^2 trF_1^2 + c_8 trR^2 trF_2^2 + c_9\ trF_1^2 trF_2^2 + c_{10} trR^4]$$

Here we simply use the method of expanding and comparing that we have already exploited to calculate the factorization of low dimensional groups.
We set $c_1 = 1$ and proceed to obtain $c_4 = \frac{1}{1536(2\pi)^5}$, $c_{10} = \frac{1}{384(2\pi)^5}$, $c_2 = -\frac{2}{15}$, $c_3 = -\frac{2}{15}$ and $c_5 = \frac{1}{6(2\pi)^5}$ upon comparison with, respectively, $(trR^2)^3$, $trR^4 trR^2$, $trF_1^2 trR^4$, $trF_2^2 trR^4$ and $\left(trF_1^2\right)^3$.
These newly calculated coefficients allow us to tackle $c_7$. In fact, $c_7$ is represented by an overdetermined system of equations, along the lines of the systems found in section 2.5. Specifically, two equations describe $c_7$: one is obtained comparing the terms proportional to $(trR^2)^2 trF_1^2$, the other comparing those proportional to $trR^2\left(trF_1^2\right)^2$.
All in all, the system yields two different values, with the obvious conclusion that $SU(2) \times SU(2) \times U(1)^{490}$ is indeed an anomalous theory. We can think of the overdetermined system as the anomalous part of $SU(2)$. One can also check that if the calculations are repeated for $F_2$, $c_8$ will be analogously impossible to retrieve: the reason is of course the fact that $c_8$ carries the anomaly of the second $SU(2)$ theory, which is related to the curvature $F_2$ from which the overdetermined system arises.

As a corollary, the above result sets constraints to theories in 496 dimensions constructed increasing the number of $SU(2)$ theories and simultaneously decreasing the order of $U(1)$.
Let us elaborate on this thought. $R$ is a property of the theory as a combination of supergravity and Yang-Mills, therefore it does not change in the domain of theories with 496 dimensions: in fact, it will be clear in the next examples that the first term of the anomaly is always

$$I_{12}^{TOT} \propto \frac{1}{64(2\pi)^5}\left(\frac{1}{6} trR^4 trR^2 + \frac{1}{24}(trR^2)^3\right)$$

Consequently, this term is invariant under addition of $SU(2)$ theories.
As far as the curvature $F$ is concerned, adding more theories increases the number of labeled curvatures, since these depend on the individual contributions.
However, the system of equations that invalidates $c_7$ only depends on $trR^2$ and $trF_1^2$. The latter term survives upon addition of $SU(2)$ theories, since the second term of the total anomaly polynomial trivially becomes proportional to:

$$I_{12}^{TOT} \propto \left(trF_1{}^2 + trF_2{}^2 \ldots + trF_n{}^2\right)$$

and it is clear that no term $trF_i{}^2$ will cancel the pre-existing $trF_1{}^2$.
By induction, therefore, the anomaly is carried along in all groups of the form

$$G_1 = \prod_{i=1}^{n} SU(2)_i \times U(1)^{496-3n}$$

so that they are all anomalous.
Accidentally, combinations with the anomaly-free $E_8$ are not permitted as long as there is a $SU(2)$ theory, since this will carry an anomaly that cannot be cancelled upon addition of $E_8$. One way to think about this is to consider that, as far as anomalies are concerned, the proof used for $G_1$ is invariant upon permutation of $U(1)$ and $E_8$.

### 3.3 Combinations of $Sp(2)$ and $U(1)$

The simplest theory in this case is $Sp(2) \times Usp(4) \times U(1)^{476}$. The YM contribution is obtained summing two anomalies arising from $Sp(2)$ and one from $U(1)^{476}$. As usual for the latter we set $tr_{\mathcal{R}}1 = dimU(1)^{476} = 476$ and $F = 0$. As far the former is concerned, we insert the trace identities for $Sp(2)$ [[2.5.1.1], (2.5.1.2), (2.5.1.3)] in (3.2.2).

$$I_{12}^{U(1)^{476}} = \frac{1}{64(2\pi)^5}\left(\frac{476}{5670}trR^6 + \frac{476}{4320}trR^4trR^2 + \frac{476}{10368}(trR^2)^3\right) \quad (3.3.1)$$

$$\begin{aligned}
I_{12}^{Sp(2)} = &\frac{1}{64(2\pi)^5}\left(\frac{10}{5670}trR^6 + \frac{10}{4320}trR^4trR^2 + \frac{10}{10368}(trR^2)^3\right) \\
&- \frac{3}{16(2\pi)^5}(trF^2)\left(\frac{1}{360}trR^4 + \frac{1}{288}(trR^2)^2\right) \\
&+ \frac{1}{1152(2\pi)^5}trR^2\left(3(trF^2)^2 + 12trF^2\right) \\
&- \frac{1}{720(2\pi)^5}\left[-\frac{9}{2}(trF^2)^3 + 42trF^4trF^2\right]
\end{aligned} \quad (3.3.2)$$

Starting from (3.3.2) we trivially generalize the anomaly to include a second contribution of $Sp(2)$:

$$\begin{aligned}
I_{12}^{Sp(2)\times Sp(2)} = &\frac{1}{64(2\pi)^5}\left(\frac{20}{5670}trR^6 + \frac{20}{4320}trR^4trR^2 + \frac{20}{10368}(trR^2)^3\right) \\
&- \frac{3}{16(2\pi)^5}(trF_1{}^2 + trF_2{}^2)\left(\frac{1}{360}trR^4 + \frac{1}{288}(trR^2)^2\right) \\
&+ \frac{1}{1152(2\pi)^5}trR^2\left(3(trF_1{}^2)^2 + 12trF_1{}^2 + 3(trF_2{}^2)^2 + 12trF_2{}^2\right)
\end{aligned}$$

$$-\frac{1}{720(2\pi)^5}\left[-\frac{9}{2}\left(trF_1{}^2\right)^3 + 42trF_1{}^4 trF_1{}^2 - \frac{9}{2}\left(trF_2{}^2\right)^3 + 42trF_2{}^4 trF_2{}^2\right]$$

$$(3.3.3)$$

The Yang-Mills is then easily obtained from the individual contributions:

$$I_{12}^{YM} = \frac{1}{64(2\pi)^5}\left(\frac{496}{5670}trR^6 + \frac{496}{4320}trR^4 trR^2 + \frac{496}{10368}(trR^2)^3\right)$$
$$-\frac{3}{16(2\pi)^5}\left(trF_1{}^2 + trF_2{}^2\right)\left(\frac{1}{360}trR^4 + \frac{1}{288}(trR^2)^2\right)$$
$$+\frac{1}{1152(2\pi)^5}trR^2\left(3\left(trF_1{}^2\right)^2 + 12trF_1{}^2 + 3\left(trF_2{}^2\right)^2 + 12trF_2{}^2\right)$$
$$-\frac{1}{720(2\pi)^5}\left[-\frac{9}{2}\left(trF_1{}^2\right)^3 + 42trF_1{}^4 trF_1{}^2 - \frac{9}{2}\left(trF_2{}^2\right)^3 + 42trF_2{}^4 trF_2{}^2\right]$$

$$(3.3.4)$$

Finally, the total anomaly polynomial is retrieved coupling the anomalies of Yang-Mills theory and supergravity, exactly as done in section 3.2.

$$I_{12}^{TOT} = \frac{1}{64(2\pi)^5}\left(\frac{1}{6}trR^4 trR^2 + \frac{1}{24}(trR^2)^3\right)$$
$$-\frac{3}{16(2\pi)^5}\left(trF_1{}^2 + trF_2{}^2\right)\left(\frac{1}{360}trR^4 + \frac{1}{288}(trR^2)^2\right)$$
$$+\frac{1}{1152(2\pi)^5}trR^2\left(3\left(trF_1{}^2\right)^2 + 12trF_1{}^2 + 3\left(trF_2{}^2\right)^2 + 12trF_2{}^2\right)$$
$$-\frac{1}{720(2\pi)^5}\left[-\frac{9}{2}\left(trF_1{}^2\right)^3 + 42trF_1{}^4 trF_1{}^2 - \frac{9}{2}\left(trF_2{}^2\right)^3 + 42trF_2{}^4 trF_2{}^2\right]$$

$$(3.3.5)$$

It is worthwhile pointing out that, as expected, the first term is the same as (3.2.7). The anomaly polynomial must, for the usual reasons, factorize in such a way that:

$$\left(c_1 trR^2 + c_2\ trF_1{}^2 + c_3\ trF_2{}^2\right)\cdot\left[c_4\ (trR^2)^2 + c_5\left(\ trF_1{}^2\right)^2 + c_6\left(\ trF_2{}^2\right)^2\right.$$

$$\left.+c_7 trR^2 trF_1{}^2 + c_8 trR^2 trF_2{}^2 + c_9\ trF_1{}^2 trF_2{}^2 + c_{10}\ trR^4 + c_{11} trF_1{}^4 + c_{12} trF_2{}^4\right]$$

As a starting point we set $c_1 = 1$. Then we obtain $c_4 = \frac{1}{1536(2\pi)^5}$, $c_{10} = \frac{1}{384(2\pi)^5}$, $c_3 = -\frac{1}{5}$, $c_2 = -\frac{1}{5}$, $c_{11} = \frac{7}{24(2\pi)^5}$, $c_6 = -\frac{1}{32(2\pi)^5}$ and $c_5 = -\frac{1}{32(2\pi)^5}$.

These values have been obtained comparing terms proportional to $(trR^2)^3$, $trR^4 trR^2$, $trF_2{}^2 trR^4$, $trF_1{}^2 trR^4$, $trF_1{}^2 trF_1{}^4$, $\left(trF_2{}^2\right)^3$ and $\left(trF_1{}^2\right)^3$, respectively.

Such constants leave us with an usual overdetermined system of equations for $c_7$ obtained comparing $\left(trF_1{}^2\right)^2 trR^2$ and $trF_1{}^2 (trR^2)^2$.

A trivial inspection shows that the system is impossible and that therefore $Sp(2) \times Sp(2) \times U(1)^{476}$ is anomalous. In conclusion,

$$G_2 = \prod_{i=1}^{m} Sp(2)_i \times U(1)^{496-10m}$$

is found to be anomalous for the same reasons as the previous generalized group $G_1$. It is not even possible to combine $E_8$ with $Sp(2)$ in such a way to obtain an anomaly-free theory in 496 dimensions. This is quickly ruled out by the anomaly carried by $Sp(2)$, which survives upon addition of anomaly-free as well as anomalous theories.

### 3.4 Combinations of $SU(2)$, $Sp(2)$ and $U(1)$

We expect all theories of this form to be anomalous, the reason being that they are constructed out of anomalous contribution.

As a check, however, we will try to factorize $Sp(2) \times SU(2) \times U(1)^{483}$. This will allow us to make more general conclusions for the group:

$$G_3 = \prod_{i=1}^{n} SU(2)_i \times \prod_{i=1}^{m} Sp(2)_i \times U(1)^{496-10m-3n}$$

(3.4.1)

The two anomalies that add up to the anomaly of the Yang-Mills theory are:

$$I_{12}^{U(1)^{483}} = \frac{1}{64(2\pi)^5}\left(\frac{483}{5670} trR^6 + \frac{483}{4320} trR^4 trR^2 + \frac{483}{10368}(trR^2)^3\right) \quad (3.4.2)$$

$$\begin{aligned} I_{12}^{Sp(2)\times SU(2)} = &\frac{1}{64(2\pi)^5}\left(\frac{13}{5670} trR^6 + \frac{13}{4320} trR^4 trR^2 + \frac{13}{10368}(trR^2)^3\right) \\ &- \frac{1}{16(2\pi)^5}\left(2trF_1^2 + 3trF_2^2\right)\left(\frac{1}{360} trR^4 + \frac{1}{288}(trR^2)^2\right) \\ &+ \frac{1}{1152(2\pi)^5} trR^2\left(8\left(trF_1^2\right)^2 + 3\left(trF_2^2\right)^2 + 12trF_2^2\right) \\ &- \frac{1}{720(2\pi)^5}\left[16\left(trF_1^2\right)^3 - \frac{9}{2}\left(trF_2^2\right)^3 + 42trF_2^4 trF_2^2\right] \end{aligned}$$

(3.4.3)

So that the anomaly of the Yang-Mills is:

$$\begin{aligned} I_{12}^{YM} = &\frac{1}{64(2\pi)^5}\left(\frac{496}{5670} trR^6 + \frac{496}{4320} trR^4 trR^2 + \frac{496}{10368}(trR^2)^3\right) \\ &- \frac{1}{16(2\pi)^5}\left(2trF_1^2 + 3trF_2^2\right)\left(\frac{1}{360} trR^4 + \frac{1}{288}(trR^2)^2\right) \\ &+ \frac{1}{1152(2\pi)^5} trR^2\left(8\left(trF_1^2\right)^2 + 3\left(trF_2^2\right)^2 + 12trF_2^2\right) \\ &- \frac{1}{720(2\pi)^5}\left[16\left(trF_1^2\right)^3 - \frac{9}{2}\left(trF_2^2\right)^3 + 42trF_2^4 trF_2^2\right] \end{aligned}$$

(3.4.4)

and finally we retrieve the total anomaly polynomial:

$$
\begin{aligned}
I_{12}^{TOT} = \ & \frac{1}{64(2\pi)^5}\left(\frac{1}{6}trR^4 trR^2 + \frac{1}{24}(trR^2)^3\right) \\
& - \frac{1}{16(2\pi)^5}\left(2trF_1{}^2 + 3trF_2{}^2\right)\left(\frac{1}{360}trR^4 + \frac{1}{288}(trR^2)^2\right) \\
& + \frac{1}{1152(2\pi)^5}trR^2\left(8\left(trF_1{}^2\right)^2 + 3\left(trF_2{}^2\right)^2 + 12trF_2{}^2\right) \\
& - \frac{1}{720(2\pi)^5}\left[16\left(trF_1{}^2\right)^3 - \frac{9}{2}\left(trF_2{}^2\right)^3 + 42trF_2{}^4 trF_2{}^2\right]
\end{aligned}
\tag{3.4.5}
$$

The proposed factorization is:

$$
\left(c_1 trR^2 + c_2\, trF_1{}^2 + c_3\, trF_2{}^2\right)\cdot[c_4(trR^2)^2 + c_5\left(trF_1{}^2\right)^2 + c_6\left(trF_2{}^2\right)^2
$$

$$
+ c_7\, trR^2 trF_1{}^2 + c_8\, trR^2 trF_2{}^2 + c_9\, trF_1{}^2 trF_2{}^2 + c_{10}\, trR^4 + c_{11}\, trF_2{}^4]
$$

However, following the same method developed above and comparing the usual terms in $F_1$, we check that the overdetermined system of equations is exactly the one found for $SU(2)\times SU(2)\times U(1)^{490}$. Comparing the terms in $F_2$, we instead obtain the same equations of $Sp(2)\times Sp(2)\times U(1)^{476}$. This reinforces our ideas that the overdetermined systems of equations carry the information of the anomaly for the group out of which they are constructed.

Either anomaly is not cancelled upon addition of any term. and $G_3$ is anomalous.

This last section concludes our analysis of theories in D=10. It is found that, indeed, $SO(32)$, $E_8\times E_8$, $U(1)^{496}$ and $E_8\times U(1)^{248}$ are the only groups where the anomaly is cancelled.

However, the hope is that this explicit analysis could be of any help in understanding the quite vast literature in anomaly cancellation. We also hope to have conveyed a convincing argument on how anomalies behave when theories are considered individually and in combination with other ones.

## Acknowledgements

First and foremost I would like to thank professor Yuji Tachikawa, without whose advice and daily support this project would have never started nor it could have ever been completed.

I would also like to thank the University of Tokyo for the opportunity to participate in the UTRIP program for undergraduate students and for funding my six-week research stay in Japan.

Last but not least, I would like to thank all the professors, postdocs and graduate students of the Department of Physics of the University of Tokyo, who have made me feel at home for the entire duration of the program.

# Appendix

## A. Constructing a Green-Schwarz term for $U(1)^{496}$

The anomaly of the YM theory coincides with the anomaly arising from $U(1)$.

$$I_{12}^{YM} = I_{12}^{U(1)^{496}} = \frac{1}{64(2\pi)^5}\left(\frac{496}{5670}trR^6 + \frac{496}{4320}trR^4trR^2 + \frac{496}{10368}(trR^2)^3\right)$$

The total anomaly becomes:

$$I_{12}^{TOT} = \frac{1}{64(2\pi)^5}\left(\frac{1}{6}trR^4trR^2 + \frac{1}{24}(trR^2)^3\right)$$

where we have set, as usual, $F = 0$. From the total anomaly polynomial we can trivially see by inspection that the term $trR^2$ is common to both addends. Therefore, the factorization of the polynomial is readily obtained:

$$I_{12}^{TOT} = \frac{1}{64(2\pi)^5}trR^2\left(\frac{1}{6}trR^4 + \frac{1}{24}(trR^2)^2\right)$$

The counterterm needed to cancel the polynomial has the following form:

$$\Delta\hat{I}_{12} = -trR^2 \wedge X_8$$

since, of course, , $F = 0$ for the theory we are considering. The Green-Schwarz term is chosen in such a way that it cancels the anomaly. Specifically:

$$X_8 = \frac{1}{64(2\pi)^5}\left(\frac{1}{6}trR^4 + \frac{1}{24}(trR^2)^2\right)$$

so that

$$\Delta\hat{I}_{12} = -\frac{1}{64(2\pi)^5}trR^2\left(\frac{1}{6}trR^4 + \frac{1}{24}(trR^2)^2\right) = -I_{12}^{TOT}$$

$$I_{12}^{TOT} + \Delta\hat{I}_{12} = 0$$

e.g. the total anomaly polynomial is cancelled via the Green-Schwarz Mechanism.

## B. Constructing a Green-Schwarz term for $E_8 \times U(1)^{248}$

In order to construct the anomaly of the YM theory we need the individual contributions from $E_8$ and $U(1)^{248}$:

$$I_{12}^{U(1)^{248}} = \frac{1}{64(2\pi)^5}\left(\frac{248}{5670}trR^6 + \frac{248}{4320}trR^4trR^2 + \frac{248}{10368}(trR^2)^3\right)$$

For $E_8$ we use the following trace identities [3]:

$$TrF^2 = trF^2$$

$$TrF^4 = \frac{1}{100}(trF^2)^2$$

$$TrF^6 = \frac{1}{7200}(trF^2)^3$$

Then the contribution to the anomaly is:

$$I_{12}^{E_8} = \frac{1}{64(2\pi)^5}\left(\frac{248}{5670}trR^6 + \frac{248}{4320}trR^4trR^2 + \frac{248}{10368}(trR^2)^3\right)$$
$$- \frac{1}{32(2\pi)^5}(trF^2)\left(\frac{1}{360}trR^4 + \frac{1}{288}(trR^2)^2\right) + \frac{1}{115200(2\pi)^5}(trF^2)^2trR^2$$
$$- \frac{1}{518400(2\pi)^5}(trF^2)^3$$

So that the total anomaly polynomial is:

$$I_{12}^{TOT} = \frac{1}{64(2\pi)^5}\left(\frac{1}{6}trR^4trR^2 + \frac{1}{24}(trR^2)^3\right)$$
$$- \frac{1}{32(2\pi)^5}(trF^2)\left(\frac{1}{360}trR^4 + \frac{1}{288}(trR^2)^2\right) + \frac{1}{115200(2\pi)^5}(trF^2)^2trR^2$$
$$- \frac{1}{518400(2\pi)^5}(trF^2)^3$$

The factorization of the theory has the following form:

$$I_{12}^{TOT} = (c_1 trR^2 + c_2\,trF^2)\cdot[c_3(trR^2)^2 + c_4(\,trF^2)^2 + c_5\,trR^2trF^2 + c_6\,trR^4]$$

Upon expansion and comparison with the terms of the total anomaly polynomial the following coefficients are retrieved:

$$I_{12}^{TOT} = \frac{1}{384(2\pi)^5}\left(trR^2 - \frac{1}{30}trF^2\right)\left[\frac{(trR^2)^2}{4} + \frac{(\,trF^2)^2}{450} - \frac{trR^2trF^2}{30} + \,trR^4\right]$$

We have obtained the factorized form that allows us cancelling the anomaly via the Green-Schwarz Mechanism. The particular gauge term is:

$$X_8 = \frac{1}{384(2\pi)^5} \left[ \frac{(trR^2)^2}{4} + \frac{(trF^2)^2}{450} - \frac{trR^2 trF^2}{30} + trR^4 \right]$$

So that:

$$\Delta \hat{I}_{12} = \frac{1}{384(2\pi)^5} \left( \frac{1}{30} trF^2 - trR^2 \right) \left[ \frac{(trR^2)^2}{4} + \frac{(trF^2)^2}{450} - \frac{trR^2 trF^2}{30} + trR^4 \right]$$

As required to cancel the anomaly via the Green-Schwarz mechanism.